\documentclass[aoas,preprint]{imsart}\usepackage[]{graphicx}\usepackage[]{color}
\makeatletter
\def\maxwidth{ %
  \ifdim\Gin@nat@width>\linewidth
    \linewidth
  \else
    \Gin@nat@width
  \fi
}
\makeatother

\definecolor{fgcolor}{rgb}{0.345, 0.345, 0.345}

\usepackage{framed}
\makeatletter
 {\par\unskip\endMakeFramed%
 \at@end@of@kframe}
\makeatother

\definecolor{shadecolor}{rgb}{.97, .97, .97}
\definecolor{messagecolor}{rgb}{0, 0, 0}
\definecolor{warningcolor}{rgb}{1, 0, 1}
\definecolor{errorcolor}{rgb}{1, 0, 0}
\newenvironment{knitrout}{}{} 

\usepackage{alltt}

\usepackage{xspace}
\RequirePackage[OT1]{fontenc}
\usepackage{amsthm,amsmath,amsfonts}
\RequirePackage{natbib}
\RequirePackage[colorlinks,citecolor=blue,urlcolor=blue]{hyperref}

\newcommand{\SH}{Stehl\'ik and Hermann\xspace}
\newcommand{\IGMM}{\texttt{IGMM}\xspace}

\defcitealias{fExtremes13}{Wuertz et al., 2013}

\usepackage{geometry}
\usepackage{breakurl}
\usepackage{subfig}

\DeclareGraphicsExtensions{.pdf,.png,.jpg}

\newtheorem{theorem}{Theorem}
\newtheorem{definition}{Definition}[theorem]

\arxiv{arXiv:0000.0000}

\startlocaldefs

\endlocaldefs
\IfFileExists{upquote.sty}{\usepackage{upquote}}{}
\begin{document}

\begin{frontmatter}
\title{Rebuttal of the ``Letter to the Editor'' \\ of Annals of Applied Statistics on \\ Lambert W $\times$ F distributions and the IGMM algorithm}
\runtitle{Rebuttal of \SH}

\begin{aug}
\author{\fnms{Georg M.} \snm{Goerg}\thanksref{t1}\ead[label=e1]{im@gmge.org}}
\runauthor{G.\ M.\ Goerg}
\thankstext{t1}{This work was completed while the author was at Google.  However, the content of this work is not related to Google in any way.}
\affiliation{Google Inc.}

\address{111 8th Avenue, Google\\
         New York, NY 10011 \\
         USA \\
\printead{e1}\\
\phantom{E-mail:\ }}

\end{aug}

\begin{abstract}
I discuss comments and claims made in \citet{StehlikHermann15_LetterAoas} about skewed Lambert W $\times$ F random variables and the IGMM algorithm.  I clarify  misunderstandings about the definition and use of Lambert W $\times$ F distributions and show that most of their empirical results cannot be reproduced.  I also introduce a variant of location-scale Lambert W $\times$ F distributions that are well-defined for random variables $X \sim F$ with non-finite mean and variance.
\end{abstract}



\end{frontmatter}

\section{Introduction}

In their ``Letter to the Editor'' \citet{StehlikHermann15_LetterAoas} relate work on the exact distribution of the likelihood ratio test statistic involving the Lambert W function \citep{Stehlik03_LambertWLoglikTest} to the transformation-based approach of Lambert W $\times$ F distributions to model asymmetric data \citep{GMGLambertW_Skewed}.  See recent work by \citet{Witkovskyetal15} on logarithmic Lambert W $\times$ $\chi^2$ distributions that elegantly links these two areas of research.

\SH present a convergence analysis of the IGMM algorithm when applied to data with non-existing mean or variance.  They aim to show that using IGMM for the LATAM log-return series in \citet{GMGLambertW_Skewed} was inappropriate since their statistical analysis implies that the LATAM series does not have a finite mean. See Figure \ref{fig:plot-latam} for a time series and density plot of this dataset.\\

\begin{knitrout}
\definecolor{shadecolor}{rgb}{0.969, 0.969, 0.969}\color{fgcolor}\begin{figure}[!t]

{\centering \includegraphics[width=0.9\textwidth]{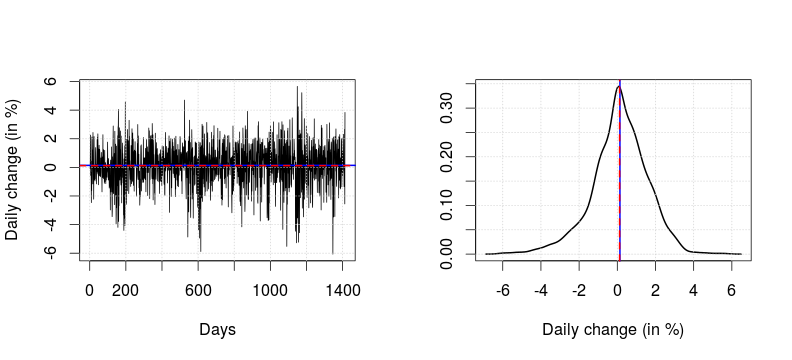} 

}

\caption[Time series plot and kernel density estimate of the LATAM daily log-return series (dataset \texttt{equityFunds} in R package \textbf{fEcofin})]{Time series plot and kernel density estimate of the LATAM daily log-return series (dataset \texttt{equityFunds} in R package \textbf{fEcofin}). Blue, solid line represents the median; red, dashed line the sample average.}\label{fig:plot-latam}
\end{figure}

\end{knitrout}

I appreciate that \SH took the time and effort to explore skewed Lambert W $\times$ F distributions and properties of the IGMM estimator. 
Here I show though that most of their findings are spurious: their methodological concerns foot on a misunderstanding of the definition of location-scale Lambert W $\times$ F distributions and their empirical findings are not reproducible.  For the remainder of this work I closely follow the organization of \citet{StehlikHermann15_LetterAoas} and address comments and claims in similar order.
When I use ``they'' or ``their'' without reference, I refer to their ``Letter to the Editor''.

\section{Heavy Tails: On three regimes of the IGMM-algorithm}

There seems to be a confusion about parameters $\mu_X$ and $\sigma_X$ of location-scale Lambert W $\times$ F distributions: they \emph{are} the expectation and standard deviation of the input random variable $X$, which means that they must exist. If they do not, then the Lambert W $\times$ F random variable is not of location-scale type from Definition 2.3 in \citet{GMGLambertW_Skewed}.  
For example, for a student-$t$ input distribution the degrees of freedom parameter $\nu$ must be greater than $2$.  This $\nu > 2$ restriction is explicitly stated in \citet{GMGLambertW_Skewed} (paragraphs above ``Notation 2.4'').  Also  \texttt{beta2tau()} in the \href{https://cran.r-project.org/web/packages/LambertW/}{\textbf{LambertW}} R package \citep{LambertWR} throws an error if $\nu \leq 2$.\footnote{Prior to \href{https://cran.r-project.org/web/packages/LambertW/}{\textbf{LambertW}} v0.5.0 \texttt{beta2tau} issued a warning to notify users about this improper application.}  Hence estimating $\mu_X$ and $\sigma_X$ is only a well-defined task for input $X$ with finite mean $\mu_X$ and variance $\sigma_X^2$.  In this case the \IGMM algorithm estimates $\mu_X$ and $\sigma_X$ by sample mean and sample standard deviation in step 8 of Algorithm 3 of \citet{GMGLambertW_Skewed}.  In practice, of course, researchers should first check if the dataset at hand satisfies this assumption (see also Section \ref{sec:pareto-tail}).\\

As a consequence, \SH's simulation study (p.\ 3, steps 1.\ - 4.) and the associated results in Table 1 are not set up correctly.  In particular, the transformation in step 2.\ is \emph{not} a location-scale Lambert W $\times$ F random variable as defined in \citet{GMGLambertW_Skewed}:  $U$ must have zero mean and unit variance, and $\mu$ and $\sigma$ must be the mean and standard deviation of $X$.  In \SH's setup the parameters do not play that role: they use $\sigma_X$ as the scale parameter, but $s$ is the scale parameter of a student-t with a standard deviation of $\sigma_X = s \cdot \sqrt{\frac{\nu}{\nu-2}}$.  For $\nu = 5$, both are well-defined with $\sigma_X = s \cdot \sqrt{\frac{5}{5-2}} \approx s \cdot 1.29$ (thus the ratio $\sigma_X / \sigma$ in their Table 1 lies approximately at $1.25$ not around $1.0$); for $\nu \in \lbrace 1, 1.5 \rbrace$ $\sigma_X$ is not finite, thus location-scale Lambert W $\times$ F inference -- with \IGMM, maximum likelihood estimation (MLE), or any other estimator -- is not well defined.

\subsubsection{New variant of location-scale Lambert W $\times$ F distributions}
\label{sec:new_location_scale}
In hindsight it would have been more clear to refer to the original transformation with $\mu_X$ and $\sigma_X$ in \citet{GMGLambertW_Skewed} as \emph{mean-variance} Lambert W $\times$ F distributions with \emph{location-scale} input in order to avoid any such confusion that arose in \SH's letter.  I will take this opportunity to introduce a new variant of location-scale Lambert W $\times$ F distributions.

\setcounter{theorem}{2}
\begin{definition}[Unrestricted Location-scale Lambert W $\times$ F]
\label{def:unrestricted-location-scale}
Let $X \sim F_X(x \mid \boldsymbol \beta)$ be a continuous location-scale random variable with location and scale parameters $c$ and $s$, and $\boldsymbol \beta$ parametrizes the distribution.  Let
\begin{align}
\label{eq:center-scale-def}
Y & =  \left( \frac{X-c}{s} \cdot e^{\gamma \frac{X-c}{s}} \right) \cdot s + c = \left (U \cdot e^{\gamma U} \right) \cdot s +  c,
\end{align}
where $U = \frac{X-c}{s}$. Then $Y$ has an \emph{unrestricted} \emph{location-scale} Lambert W $\times$ F distribution.
\end{definition}

Using general location and scale, rather than mean and standard deviation, is more natural when viewed solely as a distribution as it does not require the existence of first and second moments; e.g., Eq.\ \eqref{eq:center-scale-def} is well defined for $t$ distributions with $\nu \leq 2$.  Viewed as a data-generating process, however, transformation \eqref{eq:center-scale-def} has the main disadvantage that the image $y_i$ of a specific realization $x_i \in \mathbb{R}$ and fixed $\gamma$ depends on the value of the scale parameter of $F_X$.  Reversely, for a fixed observed $y_i$ and given variance and skewness of $Y$, the skewness parameter will be closer to zero for $F_X^{(1)}$ than for an alternative distribution $F_X^{(2)}$ if $s^{(1)} < s^{(2)}$ -- hence making $\gamma$ estimates incomparable across distributions with different scale parametrization (see Table \ref{tab:mle-ls-mv} for illustration of this property on the LATAM data). It is for this reason that I had originally defined location-scale Lambert W $\times$ F distributions only for location-scale input random variables with finite mean and variance.\\

For future reference, researchers should pick the version that is appropriate for their analysis and state whether they use the \emph{mean-variance} (original \emph{location-scale}) or the unrestricted \emph{location-scale} from Definition \ref{def:unrestricted-location-scale}.  Of course, in the latter case using the IGMM algorithm in its original form only makes sense for distributions where location and scale parameters coincide with first and/or second moments, e.g., Normal or exponential.

\begin{table}[!tb]
\begin{minipage}{.45\textwidth}
\centering
\begin{tabular}{rrrr}
  \hline
 &  Estimate &  Std. Error &  t value \\ 
  \hline
c & 0.197 & 0.037 & 5.269 \\ 
  s & 1.241 & 0.042 & 29.816 \\ 
  $\nu$ & 7.092 & 1.379 & 5.142 \\ 
  $\gamma$ & -0.053 & 0.014 & -3.944 \\ 
   \hline
\end{tabular}

\captionof{table}{mean-variance Lambert W $\times$ $t$}
\end{minipage}
\begin{minipage}{.45\textwidth}
\centering
\begin{tabular}{rrrr}
  \hline
 &  Estimate &  Std. Error &  t value \\ 
  \hline
c & 0.197 & 0.037 & 5.269 \\ 
  s & 1.241 & 0.042 & 29.814 \\ 
  $\nu$ & 7.092 & 1.379 & 5.142 \\ 
  $\gamma$ & -0.045 & 0.011 & -3.999 \\ 
   \hline
\end{tabular}

\captionof{table}{unrestricted location-scale Lambert W $\times$ $t$}
\end{minipage}
\caption{\label{tab:mle-ls-mv} Lambert W $\times$ t MLE for the LATAM data: unrestricted location-scale version has a slightly smaller $\widehat{\gamma}$ than the mean-variance version, since $\widehat{s} = 1.24 < 1.46 = \widehat{\sigma}_X$, and the non-linearity of transformation \eqref{eq:center-scale-def}.}
\end{table}

\subsection{Robust testing for normality against Pareto tail}
\label{sec:pareto-tail}

\begin{figure}[!t]

{\centering \includegraphics[width=0.9\textwidth]{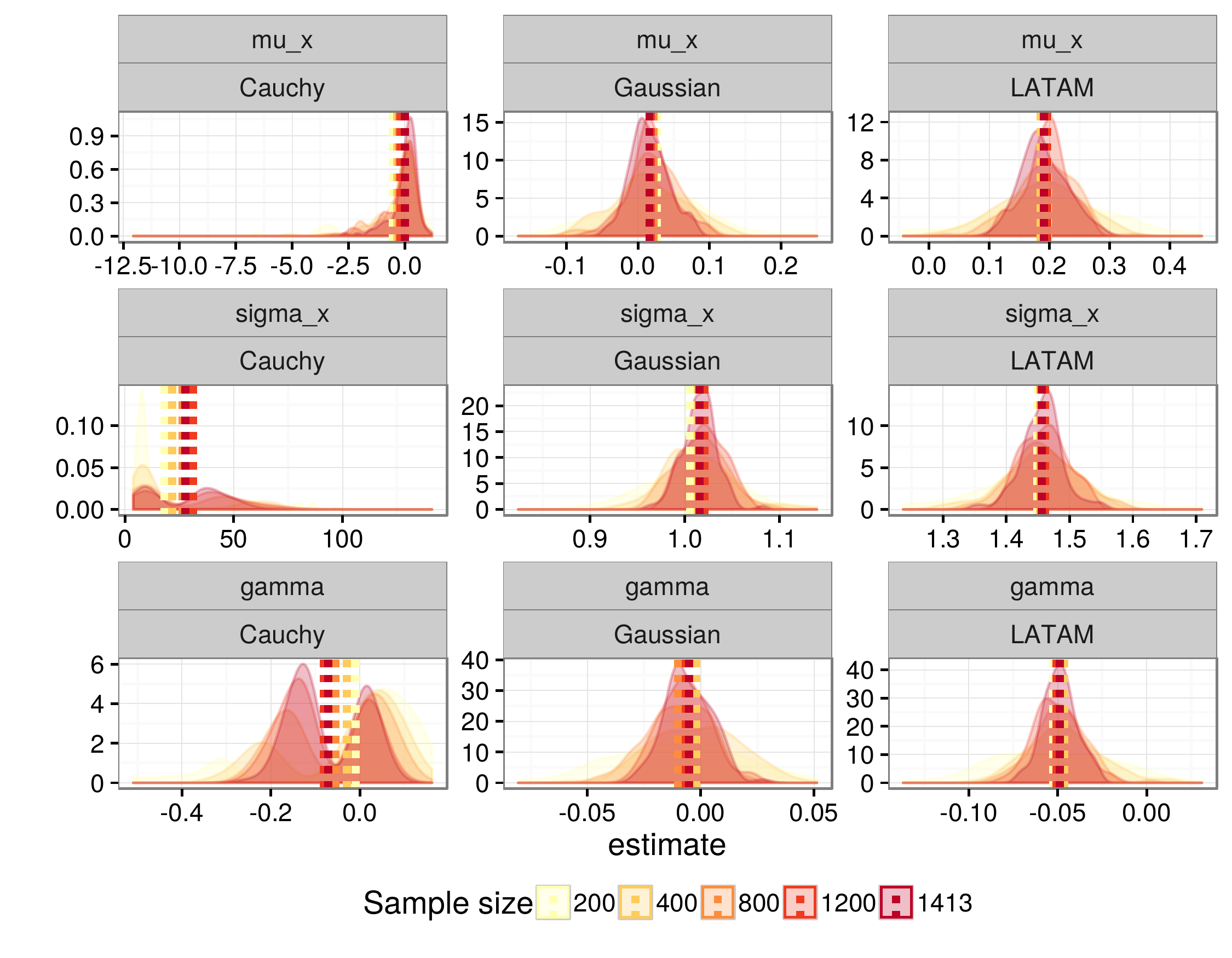} 

}

\caption[Bootstrap IGMM estimates with varying sample sizes (original size]{Bootstrap IGMM estimates with varying sample sizes (original size: $N =1413$) for simulated Cauchy and Gaussian sample and the LATAM data (vertical lines represent averages).}\label{fig:igmm-resample}
\end{figure}

Here \SH aimed to find out empirically for what type of data \IGMM yields proper statistical inference.  This is indeed an important question that I had not addressed in the original paper.  Before going into details of their analysis, I present a bootstrap approach to check if \IGMM estimates have good properties for inference on \emph{mean-variance} Lambert W $\times$ F distribution. This bootstrap convergence analysis is now readily available in the \href{https://cran.r-project.org/web/packages/LambertW/}{\textbf{LambertW}} R package \citep{LambertWR}.\\

\begin{figure}[!t]

{\centering \includegraphics[width=0.32\textwidth]{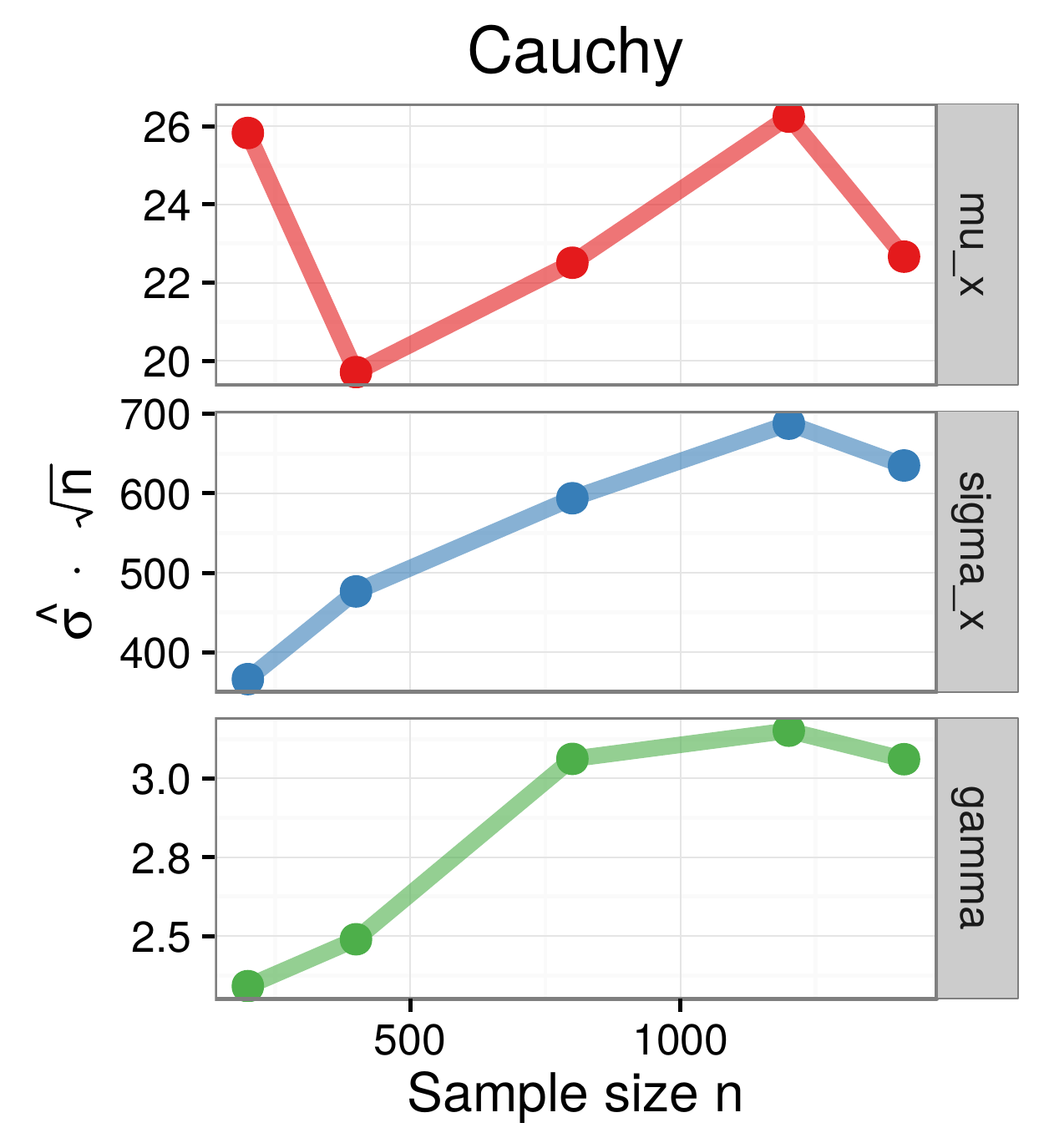} 
\includegraphics[width=0.32\textwidth]{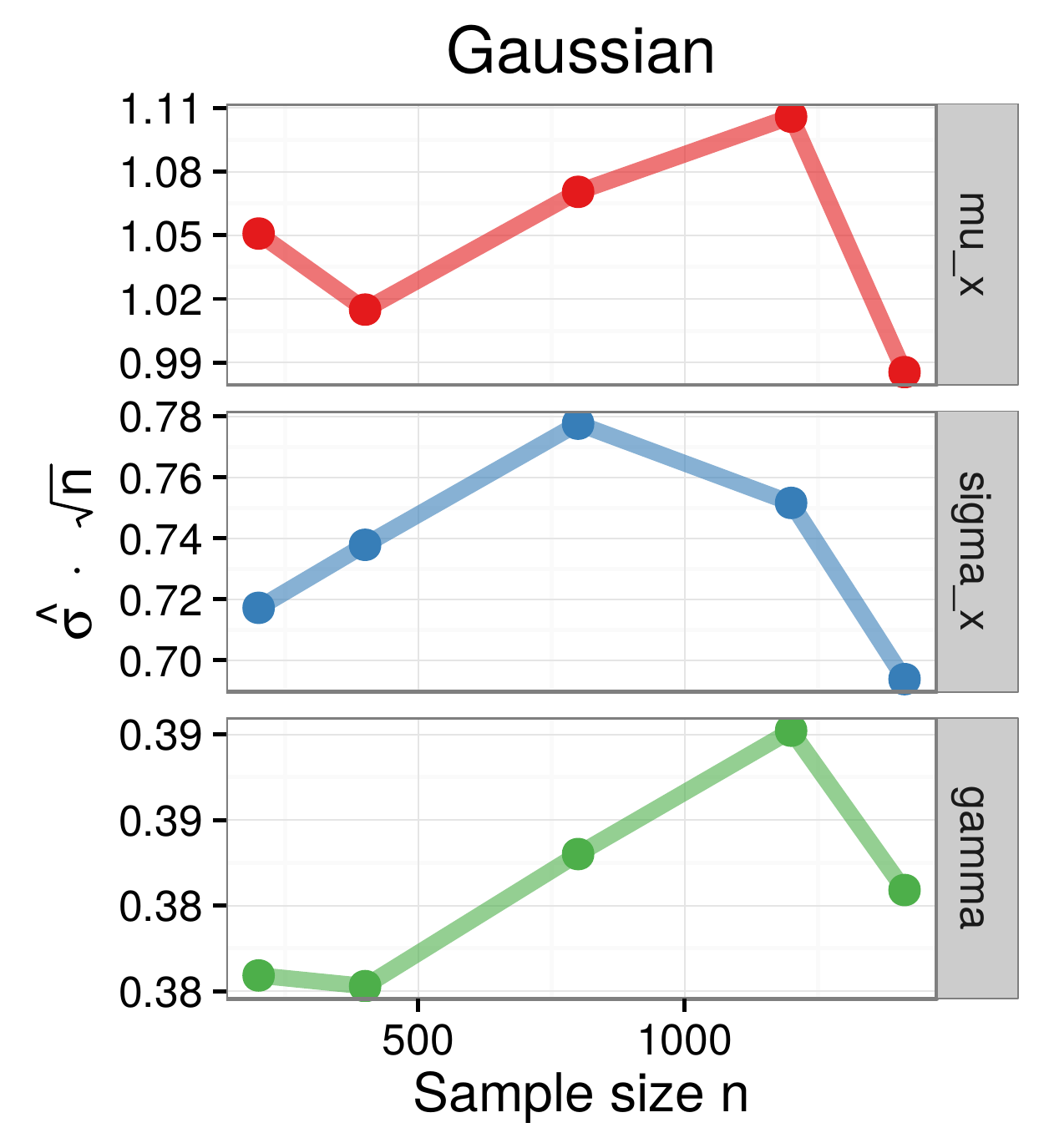} 
\includegraphics[width=0.32\textwidth]{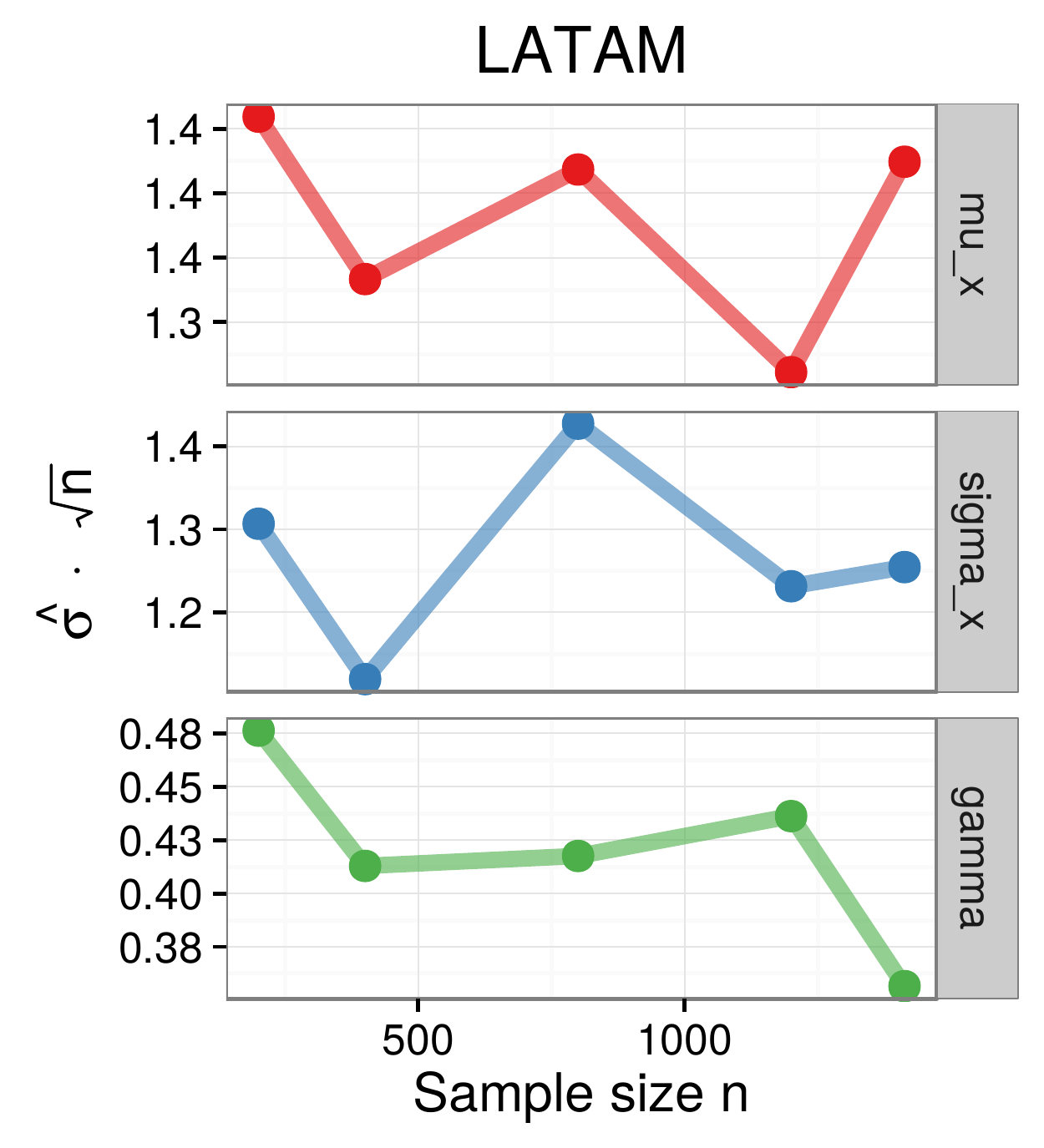} 

}

\caption[Sample standard deviation of IGMM bootstrap estimates times ]{Sample standard deviation of IGMM bootstrap estimates times $\sqrt{n}$ with 100 replications for Caucy, Gaussian, and LATAM data.}\label{fig:igmm-sd-sqrt}
\end{figure}

Figure \ref{fig:igmm-resample} shows IGMM estimates of $\tau = (\mu_X, \sigma_X, \gamma)$ for the LATAM data and for a simulated Gaussian (Regime I) and Cauchy (Regime III) sample of the same length ($N = 1413$).  I resampled the data with replacement and obtained $\widehat{\tau}_{IGMM}^{(n)}$  for varying sample sizes $n$.  As \SH show on p.\ 4, the \IGMM estimate for $\sigma_X$ (and $\mu_X$) diverge if the data lies in Regime II (or Regime III).  Figure \ref{fig:igmm-resample} compares convergence properties for increasing sample size $n$: as expected, $\widehat{\mu}_X^{(n)}$ and $\widehat{\sigma}_X^{(n)}$  do not converge for the Cauchy sample; for the Gaussian and LATAM data they do. As $n$ increases the uncertainty decreases in all estimates for the Gaussian and LATAM data, whereas the distribution of $\widehat{\mu}_X^{(n)}$ for the Cauchy sample does not change -- a well-known characteristic of averages for Cauchy samples.  Figure \ref{fig:igmm-sd-sqrt} shows that this uncertainty converges at the usual rate of $n^{-1/2}$ for a Normal sample and the LATAM data.

By contraposition of \SH's arguments we can therefore conclude that the LATAM data does not lie in Regime II or III, and thus a skewed mean-variance Lambert W $\times$ F distribution is a viable option.  One can then estimate $\tau = (\mu_X, \sigma_X, \gamma)$ with \IGMM (or any other estimator for $\tau = \tau(\theta)$, where $\theta = (\boldsymbol \beta, \gamma)$ and $\boldsymbol \beta$ parametrizes $X \sim F(x \mid \boldsymbol \beta)$).\\

\SH aimed to answer this question indirectly: they use stable distribution estimates to suggest that the LATAM data lies in Regime III, and hence can not be accurately analyzed with \IGMM. From the above bootstrap analysis we already know that it can not lie in Regime III (or II).   It is thus a natural question to ask how they arrived at conclusions supporting Regime III.  Below I discuss some shortcomings of stable distributions as a model for financial data in general, and also show that most of their results using Hill estimates cannot be reproduced.\\

\SH estimate the tail index $\alpha$ of a Pareto distribution for the input of the LATAM series based on a modified Hill estimator for dependent data \citep{Jordanova13_jordanstehlik}.   They conclude that the underlying input data falls in a regime (``Regime III'') where neither mean nor variance exist.  These findings are in contrast to the financial time series literature which is mainly concerned with the existence of finite fourth -- not first -- moments \citep{Zadrozny05, Mantegna98, Cont01_Empiricalproperties, Huismanetal01_TailIndexSmallSamples}.

To support their hypothesis about non-finite mean \SH reference empirical findings based on stable distributions in \citet{VedatBoothBruce88}.  The very same year though, \citet{AkgirayBooth88_AgainstStableDist} also published a meta analysis which rejects the stable-law model for a vast majority of $200$ stock return series and they conclude that ``[...] statistical inference should not be based on index $\alpha$ estimated from samples from stock returns''. While stable distributions have good theoretical properties as a stochastic model for financial returns, empirical evidence of finite moments has researchers led to develop less restrictive distributions \citep{KimRachevetal09, KimRachevChung06, Rosinski07_TemperedStable}.  Thus using stable distributions as an argument for non-existing mean is controversial -- as exactly these restrictions on finite moments limit their aptness as a data-generating process for stock returns \citep{Grabchak10, Lauetal90_EvidenceAgainstStable}.\\

All parametric unconditional fits considered in \citet{GMGLambertW_Skewed} (t, Lambert W $\times$ t, skew-t) can reject the hypothesis of a non-existing mean: the student-t MLE gives $\widehat{\nu} = 6.22$, with a $95 \%$ confidence interval (CI) of $(4.14, 8.3)$; the Lambert W $\times$ t fit has a similar $95 \%$ CI for $\widehat{\nu} = 7.09$: $(4.39, 9.79)$; similarly the skew-t fit gives $\widehat{\nu} = 7.16$ and skewness parameter $\widehat{\alpha} = \ensuremath{-0.8}$.  To double-check their Hill estimates of the tail parameter I use the MLE for a continuous power law fit \citep{ClausetShaliziNewman09_powerlaws} on the absolute values of the negative returns;\footnote{Estimates were obtained using the \href{http://cran.at.r-project.org/web/packages/poweRlaw/}{\textbf{poweRlaw}} R package \citep{Gillespie15}.} this yields an estimate of $\widehat{\alpha} = 3.99$  well outside Regime II or III (with optimal cutoff at $\widehat{x}_{\min} = 2.18$ -- in absolute value).  Since the LATAM series is not i.i.d., but exhibits dependence in the squared returns, I estimated a $GARCH$ model \citep{Bollerslev87} and showed that the standardized residuals also exhibit significant skewness \citep{GMGLambertW_Skewed}.  Also a $GARCH(1, 1)$  model with heavy-tailed conditional skew-t innovations remains far outside Regime II or III with $\widehat{\nu} = 9.59$.

\subsection{A graphical screening between regimes of \IGMM}

One of the main premises of the Letter is that the input of the LATAM data falls in Regime III -- as illustrated in Figure 2 of \cite{StehlikHermann15_LetterAoas}.  As shown above not only the data-driven bootstrap estimates, but also a comprehensive selection of statistical marginal and time series conditional models as well as the maximum likelihood Pareto tail estimates clearly reject their claim.

In this section I aim to replicate how \SH arrived at their conclusion; however, their findings are not reproducible.\\

\begin{figure}[!t]

{\centering \includegraphics[width=0.85\textwidth]{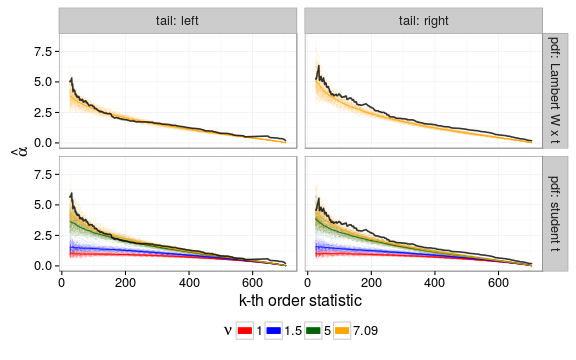} 

}

\caption[Hill curves for student t input and Lambert W ]{Hill curves for student t input and Lambert W $\times$ t output.  Simulated values (red, blue, green, orange) plus observed and transformed LATAM data (black, solid); colored solid lines are pointwise averages at each $k$. Degrees of freedom $\nu = 1$, $1.5$, and $5$ correspond to Regime III, II, and I, respectively; $\nu =  7.09 $ is the Lambert W $\times$ t MLE for $\nu$.}\label{fig:plot-hill-estimates}
\end{figure}

I follow their setup and draw i.i.d.\ samples from student-t distributions with $\nu$ = $1$, $2$, and $5$ degrees of freedom (Regime III, II, and I) and one with $\nu = 7.09$, which is the Lambert W $\times$ t MLE of $\nu$ for the LATAM data.  Furthermore, I draw random samples from the estimated \emph{true} Lambert W $\times$ t distribution with $\widehat{\theta}_{MLE} = (0.2, 1.24, 7.09, -0.05)$. 
Each simulation has the same number of samples as the LATAM data, $N = 1413$.\footnote{\SH report a different sample size of $N = 1421$.}  I then use the harmonic Hill estimator from Definition 1 of \SH with $\beta = 2$ for the simulated data and with $\beta = 1.001$ for the LATAM data.\footnote{Using the classic Hill estimator, e.g., \texttt{hillPlot()} in the \href{http://cran.at.r-project.org/web/packages/fExtremes/index.html}{\textbf{fExtremes}} R package \citepalias{fExtremes13}, is not only just marginally different in the parameter space ($\beta = 1$ vs.\ $\beta = 1.001$), but also the resulting Hill plots are essentially indistinguishable for the analyzed data and simulations.}  By taking absolute values \SH implicitly assume the series is centered around zero and left and right tail share the same properties.  Since neither assumption applies to the LATAM data, I rather estimate two Hill curves for positive and negative values separately.\footnote{Using their absolute value approach does not change the results qualitatively.} Each series was also centered by its median to ensure that Hill curves from different simulations all end at the same maximum order statistic.\\

Results from $100$ replications are shown in Figure \ref{fig:plot-hill-estimates}.  The lower panel replicates Regimes I, II, and III from Figure 2 in \citet{StehlikHermann15_LetterAoas}. However, the Hill estimates for the LATAM data \emph{cannot} be reproduced: the black, solid line does not appear below -- or even close to -- Regime III (red) as \SH show in their plots, but falls mostly on the orange Lambert W $\times$ t samples with $\nu = 7.09$, clearly \emph{above} Regime III (red). 
Overall the Hill estimates even support the left-skewed Lambert W $\times$ t distribution as a feasible marginal model for the LATAM series.

\subsection{On Regime III of IGMM}

\begin{figure}[!t]

{\centering \includegraphics[width=0.47\textwidth]{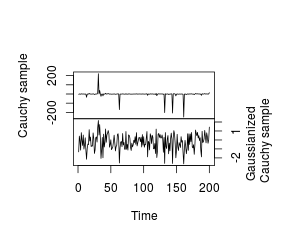} 
\includegraphics[width=0.47\textwidth]{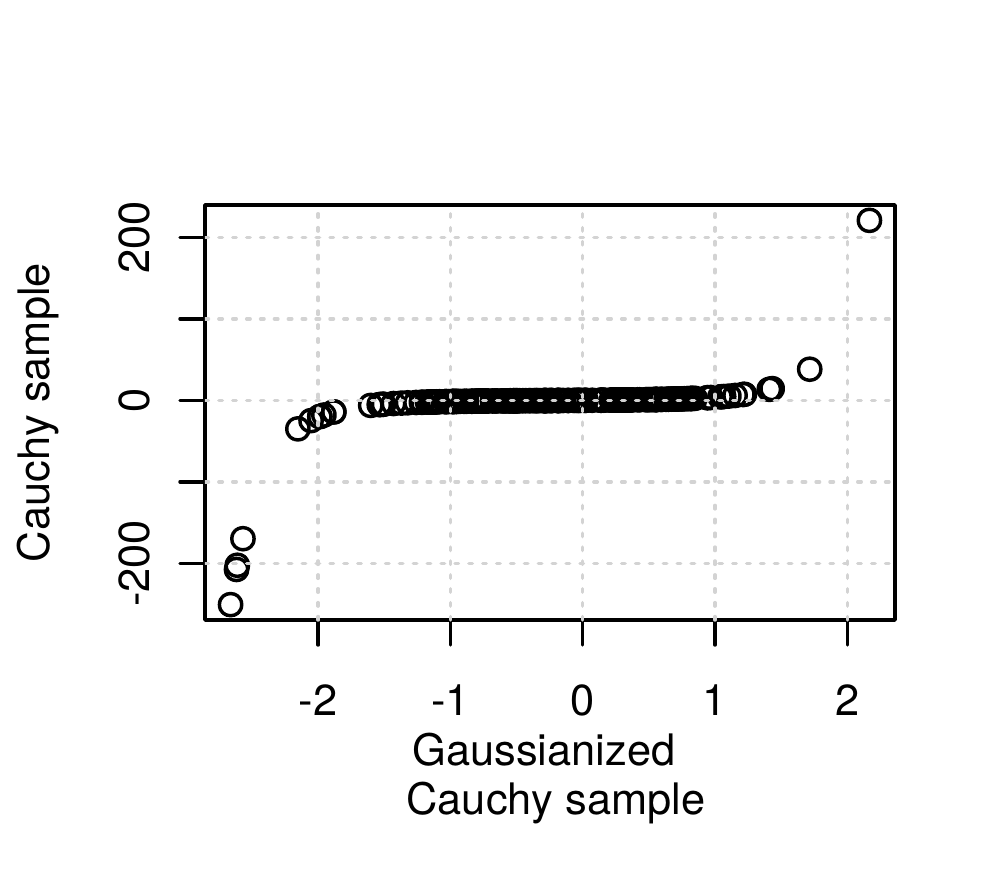} 

}

\caption{Gaussianizing a Cauchy: (left) random Cauchy sample $\mathbf{y}$ and its Gaussianized version $\mathbf{x}_{\widehat{\theta}_{IGMM}}$ using a heavy-tail Lambert W $\times$ Gaussian fit with $\widehat{\theta}_{IGMM} = (\mu_X=-0.23, \sigma_X=0.88, \delta=1.21) $ and fixed $\alpha \equiv 1$; (right) bijective mapping $\mathbf{y} \leftrightarrow \mathbf{x}_{\widehat{\theta}_{IGMM}}$.}\label{fig:cauchy-example}
\end{figure}

\SH give suggestions for a robustification of the \IGMM algorithm when $\mu_X$ and/or $\sigma_X$ do not exist.  However, in such a regime the original location-scale Lambert W $\times$ F distributions are not defined; thus estimating $\mu_X$ or $\sigma_X$ with any algorithm is not a well-posed objective after all.

For such extremely heavy-tailed data I extended skewed to heavy-tailed Lambert W $\times$ F distributions \citep{Goerg15_Gaussianize}. As an illustration, consider a random i.i.d.\ sample from a standard Cauchy distribution -- upper-left panel in Figure \ref{fig:cauchy-example}.  After estimating the parameters of the heavy-tail Lambert W $\times$ Gaussian distribution (using \IGMM with \texttt{type = "h"}) the Cauchy data can be transformed to a Gaussianized version of itself (lower-left) using a bijective mapping (right). See \citet{Goerg15_Gaussianize} for details and empirical performance of IGMM and MLE in presence of heavy tails (also covering Regime II and III cases).\\

As yet another model to test the Regime II or III hypothesis for the LATAM data I fit a two-sided heavy-tail Lambert W $\times$ Gaussian distribution via MLE: $\widehat{\delta}_l = 0.14$ 
and $\widehat{\delta}_r = 0.03$.  As expected the left tail is heavier than the right ($\widehat{\delta}_l > \widehat{\delta}_r$), and $1 / \widehat{\delta}_l = 6.92$ suggests that moments up to order $6$ exist (confidence intervals do not cover Regime \text{II or III}).\footnote{For a heavy tail Lambert W $\times$ Gaussian with $\alpha \equiv 1$ and tail parameter $\delta$ moments up to order $1 / \delta$ exist.}  Figure \ref{fig:latam-heavy-lambert1} shows that the model fits extremely well and neither parametric and non-parametric density estimates nor QQ-plots indicate major deviations from Normality for the back-transformed data (Fig.\  \ref{fig:latam-heavy-lambert2}).

\begin{knitrout}
\definecolor{shadecolor}{rgb}{0.969, 0.969, 0.969}\color{fgcolor}\begin{figure}[!t]

{\centering \subfloat[Model comparison: histogram, non-parametric kernel density estimate (KDE), a fitted Normal distribution baseline, and the parametric fit to the data.\label{fig:latam-heavy-lambert1}]{\includegraphics[width=0.5\textwidth]{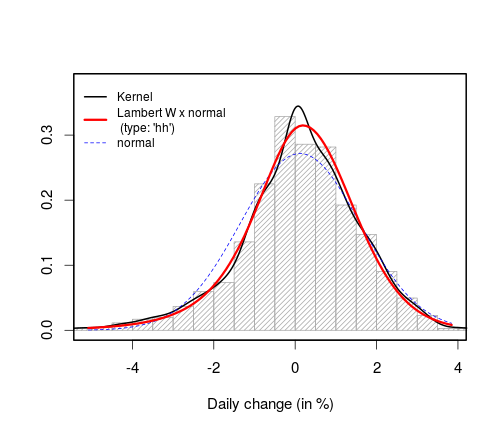} }
\subfloat[Properties of Gaussianized data: time series plot, ACF estimates, fitted Normal distribution compared to KDE, and a Normal QQ-plot.\label{fig:latam-heavy-lambert2}]{\includegraphics[width=0.45\textwidth]{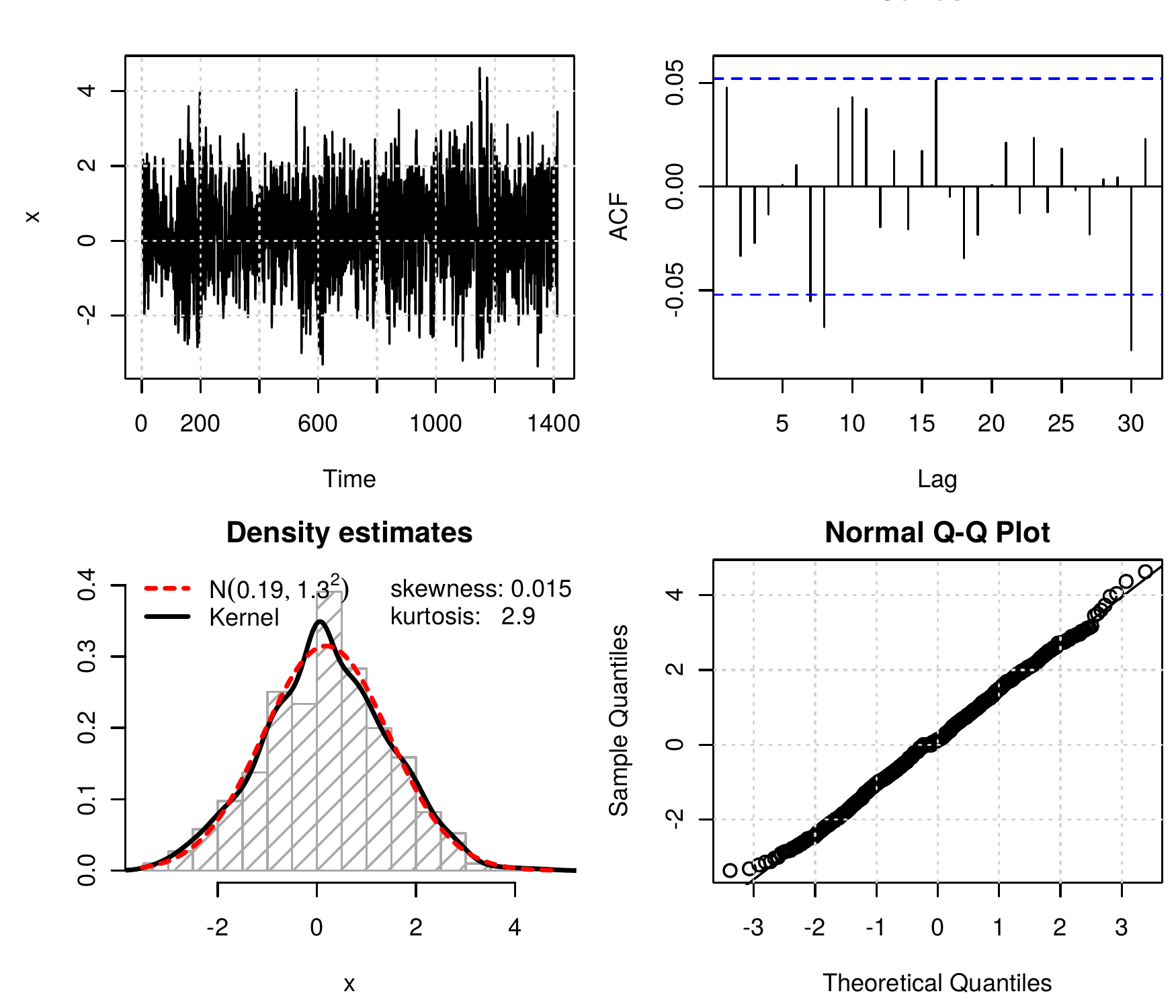} }

}

\caption[Heavy double-tail Lambert W ]{Heavy double-tail Lambert W $\times$ Gaussian distribution fit to the LATAM data.}\label{fig:latam-heavy-lambert}
\end{figure}

\end{knitrout}

\section{Skewness: On Asset returns and t-distribution}

\SH discuss a more subtle point about ``symmetrization'' since symmetry is often an inherent (physical) property of a system or object.  They opine that symmetry cannot be simply achieved by a mere variable (data) transformation.  As a popular counterexample consider the log-normal distribution.  It can either be seen as a useful distribution for non-negative right skewed data or as a transformation that achieves symmetry in the random variable / data.\\

Yet, exactly this symmetry of a system gave rise to skewed Lambert W $\times$ F distributions in the first place.  The core idea originated from observing the ``system'' of a financial market and how it processes new information into the price of a stock, currency, etc.  In principle there was no obvious reason why ``bad'' news should be more common or extreme than ``good'' news.  Put in other words, we would expect a symmetry in positive vs.\ negative returns.  However, as many empirical studies have shown \citep[see references in][]{GMGLambertW_Skewed}, financial returns are usually negatively skewed -- seemingly in conflict with the symmetry of good vs.\ bad news.  One way to accomodate the observed asymmetry is to realize that in a financial market we do not observe news \emph{per se}, but people's reaction to them.  As often is the case, people react more extremely to bad news than they do to good ones \citep[see][for related studies]{BeberBrand10, EliRao11_GoodBadNews}.

The Lambert W $\times$ F framework integrates the symmetry of news $X \sim F$ (where $F$ is a symmetric distribution with finite mean and variance), with the empirical evidence of negative skewness in financial returns, $Y$, by modeling the news processing as an asymmetric function (assuming $\mu_X = 0$ and $\sigma_X = 1$ for simplicity)
\begin{equation}
\label{eq:info-processing}
X \mapsto Y = X \cdot e^{\gamma X},
\end{equation}
where $\gamma \in \mathbb{R}$ encodes how people react to news: for $\gamma < 0$ bad news are exaggerated, for $\gamma = 0$ the quality of news has no effect, and for $\gamma > 0$ negative news affect prices less intensely.\footnote{For extremely large (in absolute value) $X$, the mapped output $Y$ is again closer to zero.  As every model, also \eqref{eq:info-processing} is an approximation and for practical purposes such extreme values can be ignored as they usually occur with miniscule probability. For details see comments on the non-principal branch probability $p_{(-1)}$ in \citet{GMGLambertW_Skewed}.}  Clearly this is only a model and I do not claim that \eqref{eq:info-processing} is the true data-generating process of financial markets.  Yet, it conveniently embeds symmetry of news, asymmetric information processing, and empirical evidence of negative skewness in financial data in a statistical model that can be estimated from observed data. 

If one cares about symmetry of a system, then one must decide if transforming random variables within the skewed Lambert W $\times$ F framework is appropriate given domain-specific context.  If inherent symmetry is not important, then one can ignore the random variable (or data) transformation and only view Lambert W $\times$ F as yet another asymmetric distribution.\\

In Section 4 they use skewed input $U$, transform it via
\begin{equation}
\label{eq:SH_trafo}
Y = \left( U \cdot \exp (-b U) \right) c + a,
\end{equation}
use \IGMM to obtain an estimate of $U$, and finally test for symmetry to show that the method fails to recognize the skewness in the input.

Again, this is based on a misunderstanding of the Lambert W $\times$ F distributions in the location-scale setting and what IGMM aims to estimate.  First, $U$ must have zero mean and unit variance, which is in general not true for the skew-t distribution. Thus \eqref{eq:SH_trafo} describes a non-central, non-scale Lambert W $\times$ skew-t distribution \citep[Definition 2.1][]{GMGLambertW_Skewed}, which is furthermore scaled (by $c$) and shifted (by $a$). This is not a location-scale Lambert W $\times$ F distribution; so there are no $\mu_X$ or $\sigma_X$ to be estimated.\footnote{According to the new Definition \ref{def:unrestricted-location-scale} from Section \ref{sec:new_location_scale} this can be now identified as an \emph{unrestricted} location-scale Lambert W $\times$ t random variable.}   Secondly, the authors assume that the estimate of $U$ from \IGMM should be close to their simulated skew-t $U$.  This
is also not the case.  By default, \IGMM sets the target skewness to $0$, thus aiming to  recover symmetric input.  For a fair comparison one should try to obtain the true $U$, in which case the target skewness must
be set to the theoretical skewness of a skew-t (or skew normal)
distribution given the respective skewness parameter.  However,
this is quite an unusual route to take as there is little value in
transforming skewed data to yet another skewed dataset.

\subsubsection{Autocorrelation}
\label{sec:autocorrelation}

\begin{figure}[!t]

{\centering \includegraphics[width=0.32\textwidth]{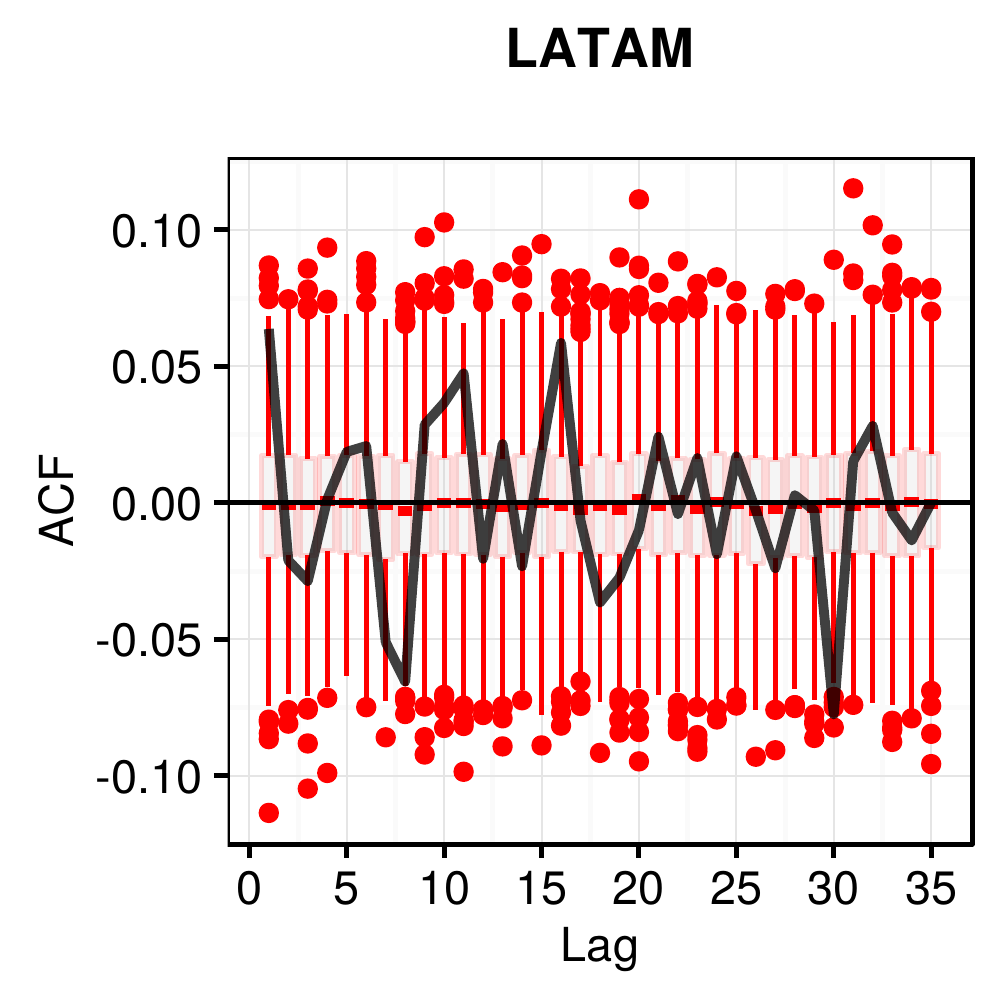} 
\includegraphics[width=0.32\textwidth]{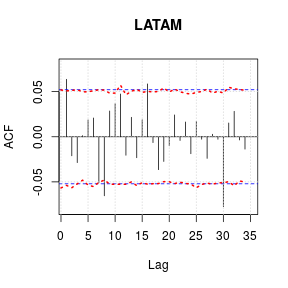} 
\includegraphics[width=0.32\textwidth]{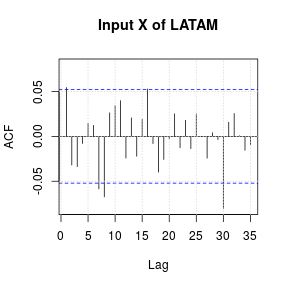} 

}

\caption[Estimated autocorrelation function (ACF) of LATAM data]{Estimated autocorrelation function (ACF) of LATAM data:  (left) bootstrapped (red, dotted)); (center) standard 95\% confidence intervals (blue, dashed); (right) ACF of back-transformed data.}\label{fig:bootstrap-acf-latam}
\end{figure}

Their findings that the ``autocorrelation function of the back-transformed data of LATAM series has observed to be significant (e.g., lags $2$, $7$, $8$, $13$, and $30$)'' are debatable: only few lags (e.g., $1$, $7$, $8$, $16$, and $30$) of the autocorrelation estimates of the LATAM series and its symmetrized input fall barely outside the $95\%$ confidence intervals (Figure \ref{fig:bootstrap-acf-latam}).  Using a multiple lag test, e.g., Box-Ljung test \citep{LjungBox78}, is inconclusive at best on rejecting white noise: testing the transformed data for all lags up to $30$ none of the $30$ p-values are below $\alpha = 0.01$; for $\alpha = 0.05$ 18 reject the null (note multiple hypothesis testing though).   In any case, the ACF plot indicates already that this fund would not be the first choice for a successful trading strategy.\\

It is not clear from their Letter why these autocorrelation findings -- even if they showed relevant deviations from white noise -- matter for the main points of the paper.  In Section 7.2, p.\ 29 of \citet{GMGLambertW_Skewed} I only considered the unconditional distribution for the sake of illustrating the method.  Lambert W $\times$ F 
time series models -- including ARMA, GARCH, and SV models -- were far beyond
the scope of the paper and would not have added any further insights on the new methodology to model and symmetrize skewed data.  I thus did not elaborate on this time series model fitting exercise.

\section{Discussion}
Based on claims in \citet{StehlikHermann15_LetterAoas} I clarify definitions and properties of location-scale Lambert W $\times$ F distributions, and explain when estimating $\mu_X$ and $\sigma_X$ is a well-defined task. 
Their methodological concerns about \IGMM are mostly spurious since they result from using incorrect definitions of skewed Lambert W $\times$ F random varibales.  As a consequence I also introduce a new variant of location-scale Lambert W $\times$ F distributions which does not rely on the existence of first and second moments in $X \sim F$.

On the applied side, \SH point out that a method of moments estimators should not be used for distributions whose moments do not exist, and perform an extensive simulation study to claim that the financial log-returns analyzed in \citet{GMGLambertW_Skewed} lie in a regime of non-finite mean. Using data-driven bootstrap estimates and several parametric distribution and time series models I show that this is not the case.  I also tried to replicate \SH's findings based on Hill estimates, but failed to reproduce most of their results.

\bibliographystyle{imsart-nameyear}
\bibliography{LambertWdistribution}

\end{document}